\documentclass[twocolumn,showpacs,aps,prl]{revtex4}

\newcommand{\bk}{{\bf k}}
\newcommand{\bq}{{\bf q}}

\newcommand{\bg}{{\bf g}}

\newcommand{\br}{{\bf r}}

\newcommand{\beq}{\begin{eqnarray}}
\newcommand{\eeq}{\end{eqnarray}}
\newcommand{\beqq}{\begin{eqnarray*}}
\newcommand{\eeqq}{\end{eqnarray*}}

\newcommand{\bp}{{\bf p}}

\newcommand{\kf}{k_F}

\usepackage{graphicx}
\usepackage{dcolumn}
\usepackage{bm}
\usepackage{color}
\usepackage{tabularx}

\usepackage{amsmath}

\begin{document}

\begin{titlepage}

\title{Odd-parity superconductivity in the vicinity of inversion symmetry breaking in spin-orbit-coupled systems} 
\author{Vladyslav Kozii and Liang Fu}
\affiliation{Department of Physics, Massachusetts Institute of Technology,
Cambridge, MA 02139, USA}

\begin{abstract}
We study superconductivity in spin-orbit-coupled systems in the vicinity of inversion symmetry breaking. We find that due to the presence of spin-orbit coupling, fluctuations of the incipient parity-breaking order generate an attractive pairing interaction in an odd-parity pairing channel, which competes with the $s$-wave pairing.
We show that applying a Zeeman field suppresses the $s$-wave pairing and promotes the odd-parity superconducting state.  Our work provides a new mechanism
for odd-parity pairing and opens a route to novel topological superconductivity.
\end{abstract}


\maketitle

\draft

\vspace{2mm}

\end{titlepage}

Over the last few years, the search for unconventional superconductors has received a new impetus from the study of topological phases of matter.
After early works on superfluid Helium-3~\cite{volovik} and recent developments on topological insulators,
it has been theoretically established~\cite{schynder} that superconducting states can be classified by their topological properties.
Unlike conventional $s$-wave superconductors, topological superconductors are predicted to harbor exotic quasiparticle excitations on the boundary.
There is currently intensive effort searching for topological superconductivity in naturally-occurring materials (see for example, Ref.~\cite{fuberg,ando,stroscio, wan, nagaosa, schmalian, hosur, mos2, sigrist, yang, andofu}), though definitive experimental evidence is still lacking.
For the majority of superconductors that are time-reversal and inversion symmetric,
the single most important requirement for being topologically nontrivial is that the pairing order parameter must be odd under spatial inversion~\cite{fuberg, sato},
e.g., $p$- or $f$-wave. This rekindles interest in finding odd-parity superconductors.

The parity of the pairing order parameter is tied with its spin.
In the absence of spin-orbit coupling,  odd-parity pairing is spin-triplet and vice versa.
It has long been known from studies on superfluid He~\cite{leggett} that triplet pairing can be induced by enhanced ferromagnetic spin fluctuations in the vicinity of ferromagnetic instability. This mechanism for triplet pairing, if realized in the solid-state, can lead to a topological superconductor~\cite{lee}, analogous to the topological superfluid He.

In this Letter, we propose an alternative mechanism for odd-parity pairing in the vicinity of nonmagnetic, inversion-symmetry-breaking phases in spin-orbit-coupled systems.
In the presence of spin-orbit interaction, such parity-breaking orders directly couple to electron's spin texture on the Fermi surface~\cite{fu}.
As a result, the fluctuations of an incipient parity-breaking order, which we call ``parity fluctuations'', generate an effective interaction that is strongly momentum and spin dependent.
Without assuming any special features of the Fermi surface, we show on general ground that this effective interaction is attractive in
both the $s$-wave and an odd-parity pairing channel.
Moreover, the pairing interactions in the two channels are found to be of the same order of magnitude, and in several cases, roughly equal.
We show that either Coulomb interaction or Zeeman field suppresses the $s$-wave pairing and promotes the odd-parity superconducting state on the border of parity-breaking order.
Finally, we propose 
the pyrochlore oxide Cd$_2$Re$_2$O$_7$ and doped SrTiO$_3$ heterostructures as candidate systems 
where odd-parity superconductivity meditated by parity fluctuations may be realized.

In this work, we consider parity-breaking orders that are time-reversal invariant and carry zero momentum.
Such order may originate from an unstable odd-parity phonon or the electron-electron interaction.
The order parameter can be represented by a Hermitian fermion bilinear operator ${\hat Q}$ with the same symmetry, which takes the form 
\beq
{\hat Q} &=& \sum_{\bk, \alpha \beta} \Gamma_{\alpha \beta}(\bk) c^\dagger_{\bk \alpha }  c_{\bk \beta}, \; \textrm{with } \Gamma^\dagger(\bk)= \Gamma(\bk). \label{op} 
\eeq
Here $\alpha, \beta$ are pseudospin indices for the two degenerate states at every $\bk$. In spin-orbit-coupled systems, these states are not spin eigenstates, but remain degenerate in the presence of time-reversal ($\Theta$) and inversion ($P$) symmetry~\cite{fukane}. For simplicity of notation, we have chosen an Ising-type parity-breaking order in Eq.(\ref{op}). Vector and high-rank tensor orders are described by a multiplet of Hermitian operators denoted by ${\hat Q}^\mu$; these orders will be encountered later.

Different types of parity-breaking orders are classified by their transformation properties under crystal symmetry operations, which act on electron's spin and momentum jointly.
Before proceeding to the symmetry analysis, we emphasize that the form factors for physical observables, such as $\Gamma(\bk)$ for $\hat Q$, depend on the basis for the doubly degenerate energy band. For the purpose of this work, it is most convenient to choose the ``manifestly covariant Bloch basis'' (MCBB)~\cite{fu}. In this basis, the two-component electron operator $(c^\dagger_{\bk 1}, c^\dagger_{\bk 2})$ transforms simply as a spinor field in $\bk$ space under time reversal and crystal symmetry operation $\bg \in O(3)$:
\beq
\Theta c^\dagger_{\bk \alpha} \Theta^{-1} &=&\epsilon_{\alpha \beta} c^\dagger_{-\bk \beta}  \label{tsym} \\
\bg c^\dagger_{\bk \alpha} {\bg}^{-1} &=& U_{\alpha \beta}(\bg) c^\dagger_{\bk^* \beta}, \label{bsym}
\eeq
where $\bk^* = \bg \bk$ is the star of $\bk$, $\epsilon_{\alpha \beta}$ is the Levi-Civita symbol, and $U(\bg)$ is the $U(2)$ matrix that represents the action of  $\bg$ on the pseudospin, in the same way as it acts on the spin of a free electron.

It then follows from the symmetry transformation laws (\ref{tsym}, \ref{bsym}) that the form factor  $\Gamma(\bk)$ of time-reversal-invariant and parity-breaking orders satisfies the condition
\beq
\Gamma(\bk) = \epsilon \Gamma^*(\bk) \epsilon &=& -\Gamma(-\bk),
\eeq
and hence takes the general form~\cite{fu}
\beq
\Gamma(\bk) &=& {\bf d}_\bk \cdot  {\boldsymbol \sigma},   \;  \textrm{with } {\bf d}_\bk = - {\bf d}_{-\bk} \label{gm}
\eeq
where ${\boldsymbol \sigma} = (\sigma^x, \sigma^y, \sigma^z)$ denotes Pauli matrices in pseudospin space. The $d$-vector field ${\bf d}_\bk$ defines the pseudospin splitting  in the ordered state, whose magnitude and direction vary over the Fermi surface.

It may seem counter-intuitive that nonmagnetic parity-breaking orders, such as structural distortion and orbital order, couple to electron's spin.
As we show by example in Supplementary Material~\cite{SM}, this remarkable fact is a general consequence of spin-orbit interaction in centrosymmetric systems. It will play a crucial role in meditating superconductivity in the vicinity of parity-breaking order.
In contrast, for spin-rotationally-invariant systems, the above symmetry analysis implies that
parity-breaking orders at zero momentum {\it cannot} couple directly to electrons on the Fermi surface, unlike the nematic order that is even-parity~\cite{nematic1, nematic2}.
Therefore, spin-orbit coupling is crucial for superconductivity meditated by odd-parity phonons or parity fluctuations.

In a system close to a parity-breaking instability, the effective interaction arising from the order parameter fluctuations is given by
\beq
H_{\rm eff} &=&  \sum_\bq V_\bq {\hat Q}(\bq) {\hat Q}({-\bq}) \label{eff}
\eeq
where ${\hat Q}(\bq)= {\hat Q}^\dagger(-\bq)$ is the Fourier transform of the order parameter field in real space:
\beq
{\hat Q}(\bq) =  \frac{1}{2} \sum_{\bk, \alpha \beta} (\Gamma_{\alpha \beta}(\bk+\bq) + \Gamma_{\alpha \beta}(\bk ) ) c^\dagger_{\bk+\bq\alpha }  c_{\bk \beta}  \label{qq}
\eeq
Within the random-phase approximation (RPA), $V_\bq$ can be expressed in terms of the $\bq$ dependent susceptibility: $V_\bq= I/ (1 + \chi(\bq) I)$.
$V_\bq$ is enhanced and has a maximum at $\bq=0$ close to a $\bq=0$ instability.
Restricting the effective interaction (\ref{eff}) to the Cooper pairing channel with zero total momentum, we obtain the pairing interaction
\beq
H_{\rm p} = \sum_{\bk, \bk'} V_{\alpha\beta \gamma \delta}(\bk, \bk') c^\dagger_{\bk \alpha} c^\dagger_{-\bk \beta} c_{-\bk' \gamma}c_{\bk' \delta}. \label{Hp}
\eeq
Using (\ref{gm}), (\ref{eff}) and (\ref{qq}), we find the momentum- and pseudospin-dependent interaction vertex  $V_{\alpha\beta \gamma \delta}(\bk, \bk')$ is given by
\begin{widetext}
\beq
V_{\alpha\beta \gamma \delta}(\bk, \bk')  =- \frac{1}{8} \left( V_{\bk- \bk'} ( \vec{d}_{\bk} + \vec{d}_{\bk'} ) \cdot  {\vec \sigma}_{\alpha\delta} \; ( \vec{d}_{\bk} + \vec{d}_{\bk'} ) \cdot  {\vec \sigma}_{\beta \gamma}
-V_{\bk + \bk'}( \vec{d}_{\bk} - \vec{d}_{\bk'} ) \cdot {\vec \sigma}_{\alpha\gamma} ( \vec{d}_{\bk} - \vec{d}_{\bk'} ) \cdot {\vec \sigma}_{\beta \delta}  \right). \label{vertex}
\eeq
\end{widetext}

To proceed, we expand $V_{\bk \pm \bk'}$ in the pairing interaction (\ref{vertex}) 
 in terms of spherical harmonics on the Fermi surface: $V_{\bk \pm \bk'} = V_0 \mp V_1 \hat{\bk} \cdot \hat{\bk'} + ...$.
Below we consider the leading term $V_0$. Despite $V_0$ is a constant,
the interaction vertex (\ref{vertex}) inherits the form factor of the parity-breaking order parameter $\hat Q$,  which is strongly pseudospin- and momentum-dependent.
It consists of two types of terms:
$
V= V^e + V^o
$, where 
$V^e$ contains the product of components with the same momentum:
\beq
V^e_{\alpha\beta \gamma \delta}(\bk, \bk')= -\frac{V_0 }{8}  \sum_{i,j} (d^i_{\bk} d^j_{\bk}  + d^i_{\bk'} d^j_{\bk'}) (  \sigma^i_{\alpha\delta}  \sigma^j_{\beta \gamma}  -  \sigma^i_{\alpha\gamma}   \sigma^j_{\beta \delta} ),  \nonumber
\eeq
and $V^o$ contains the cross terms:
\beq
V^o_{\alpha\beta \gamma \delta}(\bk, \bk')= - \frac{V_0 }{8}  \sum_{i,j} (d^i_{\bk} d^j_{\bk'}  + d^i_{\bk'} d^j_{\bk}) (  \sigma^i_{\alpha\delta}  \sigma^j_{\beta \gamma}  +  \sigma^i_{\alpha\gamma}   \sigma^j_{\beta \delta} ).   \nonumber
\eeq
Note that $V^e$ ($V^o$) is an even (odd) function of $\bk, \bk'$, and antisymmetric (symmetric) under exchanging the pseudospin indices either $\alpha \beta$ or $\gamma \delta$.
Therefore, $V^e$ and $V^o$ correspond to the even-parity pseudospin-singlet and odd-parity pseudospin-triplet pairing channels, respectively.

The above pairing interaction can be decomposed into different superconducting channels that belong to different representations of the crystal symmetry group.
Before proceeding, we describe the general classification of time-reversal-invariant superconducting order parameters, taking the form
\beq
{\hat F^\dagger} = \frac12 \sum_{\bk, \alpha \beta \gamma} \epsilon_{\beta \gamma} F_{\alpha \beta}(\bk)  c^\dagger_{\bk\alpha }  c^\dagger_{-\bk \gamma}.
\label{pairing}
\eeq
where the form factor $F(\bk)$ satisfies the symmetry condition
\beq
F^\dagger(\bk) = F(\bk) = - \epsilon F^*(-\bk) \epsilon.
\eeq
Moreover, it follows from (\ref{bsym}) that
 $\epsilon_{\beta \gamma} c^\dagger_{\bk\alpha }  c^\dagger_{-\bk \gamma}$
has the same transformation law  under crystal symmetry operations as $c^\dagger_{\bk\alpha }  c_{\bk \beta}$.
This implies every time-reversal-invariant superconducting order parameter has a counterpart in the particle-hole channel, with the same symmetry.
In particular, odd-parity superconducting order parameters, which have $F(\bk) = - F(-\bk)$, admit
the same classification as particle-hole orders $\hat Q$ described earlier.

To proceed with the classification, it is instructive to first consider the most symmetric group $O(3)$, the group of all joint 3D rotations and reflections
of spin and momentum, from which all point groups descend.
In this case, all possible odd-parity orders defined by the form factor (\ref{gm}) are classified by the total angular momentum $J$ and the orbital angular momentum $L$ (which must be odd)~\cite{fu}. At the lowest order $L=1$, there are three types of particle-hole orders: gyrotropic, ferroelectric and multipolar. The corresponding form factors are listed in Table I. Classification for 2D systems with $O(2)$ symmetry is also presented.

As expected from the one-to-one correspondence between particle-hole and particle-particle orders,
these form factors also classify odd-parity pairing symmetries of spin-orbit-coupled superconductors. For example, the pairing order parameter with the isotropic form factor $\Gamma_1 = \bk \cdot \boldsymbol\sigma$ coincides with a particular choice of order parameters for the Balian-Werthamer phase of He-3~\cite{footnote}. On the other hand,  the pairing order parameters with anisotropic form factors $\Gamma_2$ and $\Gamma_3$ are time-reversal invariant and spontaneously break the rotational symmetry, resulting in an odd-parity superconductor with nematic order~\cite{fu-nematic}. To our knowledge, such anisotropic phases have not been found in He-3; their existence requires spin-orbit coupling.

\begin{table}[t]
\begin{center}
\begin{tabularx}{0.5\textwidth}{| l | X |}
    \hline \text{3D system with $O(3)$ symmetry} & \text{transformation property} \\ \hline
    $\Gamma_1 (\bk) = (\hat{\bk} \cdot \boldsymbol{\sigma} )$ & \text{pseudoscalar} \\ \hline
   $ \Gamma_2^i (\bk) = [\hat{\bk} \times \boldsymbol\sigma]^i $ & \text{vector} \\ \hline
   $ \Gamma_3^{ij} (\bk) = \hat k^i \sigma^j+\hat k^j \sigma^i -\frac 23 (\hat{\bk} \cdot \boldsymbol{\sigma}) \delta^{ij}$ & \text{rank 2 tensor}\\
    \hline
\end{tabularx}
\end{center}

\begin{center}
\begin{tabularx}{0.5\textwidth}{| l | X |}
    \hline \text{2D system with $O(2)$ symmetry} & \text{transformation property} \\ \hline
    $\tilde \Gamma_1 (\bk) = \hat k^x \sigma^x + \hat k^y \sigma^y $ & \text{pseudoscalar} \\ \hline
    $\tilde\Gamma_2 (\bk) = \hat k^x \sigma^y - \hat k^y \sigma^x$ & \text{pseudoscalar}
    \\ \hline
   $ \tilde \Gamma_3^i (\bk) = \hat k^i \sigma^z $ & \text{vector} \\ \hline
   $ \tilde \Gamma_4^{ij} (\bk) =  \hat k^i \sigma^j+\hat k^j \sigma^i - (\hat{\bk} \cdot \boldsymbol{\sigma}) \delta^{ij}$ & \text{rank 2 tensor}\\
    \hline
\end{tabularx}
\end{center}

\caption{Classification of odd-parity order parameters for spin-orbit coupled systems, and their transformation properties under joint spin and momentum rotations in three and two dimensions.
Rank 2 tensor $\Gamma_3$ ($\tilde{\Gamma}_4$) is symmetric and traceless, and hence has 5 (2) independent components.
}
\label{Table1}
\end{table}

We  now use the effective interaction $H_{\rm eff}$ given by Eq.(\ref{eff}) to study superconductivity in the vicinity of each type of parity-breaking order in Table I; for multi-component operator ${\hat Q}^\mu$, summation over $\mu$ is taken.  In all cases, we restrict $H_{\rm eff}$ into Cooper pairing channel with zero momentum, and decompose the pairing interaction $H_{\rm p}$ into various superconducting channels:
\beq
H_{\rm p}=V_0 (   a_0 \hat S^\dagger \hat S + \sum_n a_n \sum_\mu {\hat F}^{\mu \dagger}_n \hat F^\mu_n ), 
\label{dec}
\eeq
where ${\hat S^\dagger} = (1/2) \sum_{\bk, \alpha \beta } \epsilon_{\alpha\beta}c^\dagger_{\bk,\alpha }  c^\dagger_{-\bk \beta}$ is the s-wave superconducting order parameter, and $F^{\mu\dagger}_n$ denotes various odd-parity order parameters defined in Eq.(\ref{pairing}) and classified in Table I. Coefficients $a_n$ take different values for different types of interactions, and all are gathered in the Table \ref{Table2}. Details of our calculation can be found in Supplementary Material~\cite{SM}. Since $V_0<0$, $a_n>0$ means attractive interaction in the corresponding pairing channel.

\begin{table}[t]
\begin{center}
\begin{tabularx}{0.5\textwidth}{| l | X | X | X | X |}
    \hline \text{type of interaction} & $a_0$ & $a_1$ & $a_2$ & $a_3$ \\ \hline
    $Q_1(\bq)Q_1(\bq)$ & 1 & 1 & $-1$& 0 \\ \hline
   $Q_2^i(\bq)Q_2^i(-\bq)$ & 2 & $-4/3$ & $1/2$ & $-1/4$  \\ \hline
 $Q_3^{ij}(\bq)Q_3^{ji}(-\bq)$ & $20/3$ & 0 & $-5/3$& $-1/2$ \\
    \hline
\end{tabularx}
\end{center}

\begin{center}
\begin{tabularx}{0.5\textwidth}{| l | X | X | X | X | X |}
    \hline \text{type of interaction} & $a_0$ & $a_1$ & $a_2$ & $a_3$ & $a_4$ \\ \hline
    $\tilde Q_1(\bq) \tilde Q_1(\bq)$ & 1 & 1 & $-1$& -1 & 0 \\ \hline
   $\tilde Q_2(\bq) \tilde Q_2(-\bq)$ & 1 & -1 & 1 & -1 & 0  \\ \hline
 $\tilde Q_3^{i}(\bq) \tilde Q_3^{i}(-\bq)$ & 1 & $-1/2$ & $-1/2$& 1 & $-1/4$
 \\ \hline
 $\tilde Q_4^{ij}(\bq) \tilde Q_4^{ji}(-\bq)$ & 4 & 0 & 0 & -4 & 0
  \\
    \hline
\end{tabularx}
\end{center}

\caption{ Decomposition of the different types of interaction into different pairing channels, see Eq. (\ref{dec}). $n=0$ denotes the $s$-wave channel;
 $n=1,...,4$ denotes the odd-parity channels classified in Table I.  $a_n>0$ corresponds to attractive pairing interaction. }
\label{Table2}
\end{table}

From Table \ref{Table2}, we obtain the superconducting instability driven by each type of parity fluctuations.
In all cases, there is an instability in the $s$-wave channel, similar to phonon-meditated pairing in conventional superconductors.
More importantly, in all cases except the multipolar orders $\Gamma_3$ and $\tilde{\Gamma}_4$,  there is also an instability in the odd-parity channel
with the same symmetry as the incipient particle-hole order that drives superconductivity.
Remarkably, for Ising type orders described by a single-component $\hat Q$, the pairing attraction in the odd-parity channel is equal to the one in the $s$-wave channel,
leading to identical superconducting transition temperatures.
For ferroelectric type orders described by a vector ${\hat Q}^i$,  the pairing attraction in the odd-parity channel is weaker than, but still of the same order of magnitude as,
 the one in the $s$-wave channel.

 Although fluctuations of multipolar orders in rotationally invariant systems do not lead to pairing in any odd-parity channel, the situation becomes different in
 real materials where the crystal symmetry is taken into account. In any crystals, the five components of rank 2 tensor $\Gamma_3$ invariably split into more than one representations of the point group. For example, for $O_h$ point group, the diagonal and off-diagonal components of $\Gamma_3$ split to form $e_g$ and $t_{2g}$ representations,
 which have $2$ and $3$ independent components respectively.  For many point groups such as $D_{4h}$, $\tilde{\Gamma}_4$ also splits into one-dimensional representations, with form factors $k_x \sigma_x - k_y \sigma_y$ and $k_x \sigma_y + k_y \sigma_x $ respectively.
We find fluctuations of such multipolar orders of reduced symmetry generate attractive pairing interaction in the odd-parity channel of the same symmetry. The interaction strength is weaker than the $s$-wave channel in the case of $e_g$ and $t_{2g}$ orders, and is equal to the latter in the case of Ising type $\tilde{\Gamma}_4$ orders~\cite{SM}.

The above finding of odd-parity pairing meditated by parity fluctuations in spin-orbit-coupled systems is the main result of this work.
It is interesting to make a comparison with the mechanism of triplet pairing meditated by spin fluctuations. In that case, the effective interaction
is given by $\sum_\bq V(\bq) {\boldsymbol s}(\bq) \cdot {\boldsymbol s}(-\bq)$, where ${\boldsymbol s}$ is the spin operator and
$V(\bq) = I/(1+I \chi^s(\bq))$ is determined by the
spin susceptibility $\chi^s(\bq)$. Importantly, to obtain the pairing interaction in the triplet channel requires $\chi^s(\bq)$ to have a nontrivial $\bq$-dependence.
Approximating $\chi^s(\bq)$ by its zeroth spherical harmonic, which is a constant, does not generate triplet pairing, simply because two electrons at the same spatial location cannot form a triplet.
In contrast, we obtained odd-parity pairing in this leading-order approximation, without relying on any special features of the susceptibility of parity-breaking order.

Given that the pairing interaction we found has comparable or even identical strengths in the $s$-wave and odd-parity channels, small residual interactions or external perturbations become important in lifting the degeneracy and eventually determine which one of the two competing pairing symmetries is realized.
An in-depth study of the effects of residual interactions necessarily involve material-specific details, which is beyond the scope of this work.
Nonetheless, it should be noted that Coulomb interaction is most repulsive and pair-breaking in the $s$-wave channel, which can make
the odd-parity pairing energetically favorable. This role of Coulomb interaction in the competition between $s$-wave and odd-parity pairings
has been recognized~\cite{fuberg} and emphasized~\cite{sau} in recent model studies.

In addition to Coulomb interaction, the $s$-wave pairing is suppressed by a magnetic field $\bf B$ that splits the spin degeneracy, which is
pair-breaking and sets the Pauli limit for the upper critical field.
However, Zeeman spin splitting has variable effects on odd-parity superconducting states in spin-orbit-coupled systems, as we show now.
First, let us consider how the doubly degenerate bands at every $\bk$, or the pseudospin, split under a Zeeman field. 
The coupling of pseudospin to Zeeman field takes the general form
\beq
H_Z =  \sum_\bk c^\dagger_{\bk } g_{ij}(\bk) B_i \sigma_j(\bk) c_{\bk}.
\eeq
The $g$-factor $g_{ij}(\bk)$ is a function of $\bk$, and can be expanded into different spherical harmonics over the Fermi surface.
Importantly, since the pseudospin operator $\sigma_j$ is defined in the manifestly covariant Bloch basis and has the same symmetry as electron's spin,
$g_{ij}(\bk)$ generally has a dominant zeroth spherical harmonic component $g^0_{ij}$.
Assuming $g_{ij}(\bk) \simeq g^0_{ij}$, we obtain a uniform spin splitting over the Fermi surface, with a spin quantization axis in the direction of $h_i = g^0_{ij} B_i$.
The Pauli limit will be absent for the odd-parity pairing if its $d$-vector ${\boldsymbol d}(\bk)$ is   perpendicular to $\bf h$, for all $\bk$ on the Fermi surface.
For example, in 2D systems with rotational symmetry, an in-plane field $B$ induces a spin splitting in the direction parallel to the field.
The odd-parity pairing with $\tilde{\Gamma}_3(\bk) = ( k_x \sigma_z, k_y \sigma_z)$, whose  $d$-vector is out of plane, is not Pauli limited.
Therefore, Zeeman field is an effective way of tuning the competition between different pairing symmetries
and promoting certain types of odd-parity superconductivity for which the Pauli limit is absent or largely enhanced.

Finally, we propose candidate materials for odd-parity superconductivity in the vicinity of parity-breaking order.
First, the pyrochlore oxide Cd$_2$Re$_2$O$_7$ undergoes a continuous parity-breaking phase transition  at $T_p=200$K~\cite{cdreo1, cdreo2} with a large mass
enhancement of conduction electrons~\cite{mass}, and becomes superconducting at $T_c=1.1$K~\cite{cdreo-sc}.
The application of high pressure has significant effects on these phases, and generates a variety of new phases identified from resistivity anomalies. Remarkably,
around a critical pressure of $P_c=4.2$GPa where the parity-breaking order is suppressed,
an anomalously large upper critical field of 7.8T is observed, which is $27$ times larger than at ambient pressure and significantly higher than the
Pauli limit 4.2T evaluated as $H_p=1.84 T_c$~\cite{pressure}. These phenomena seem to fit into the theoretical picture presented in this work. Therefore, we propose that the superconducting state of Cd$_2$Re$_2$O$_7$ around $P_c$ is
driven by parity fluctuations, and may have an odd-parity pairing symmetry.

Another  candidate system is inversion-symmetric heterostructure of doped SrTiO$_3$ with  intrinsic spin-orbit coupling~\cite{hwang}.
Bulk SrTiO$_3$ is close to the ferroelectric instability and becomes superconducting upon electron doping~\cite{behnia, marel}.
In doped SrTiO$_3$ heterostructures with the superconducting dopant layer of a few nanometers thickness,
the in-plane upper critical field exceeds the conventional Pauli limit~\cite{hwang}.
It is worthwhile to examine the possibility of an odd-parity superconducting state under a large in-plane field, as we discussed earlier.
A model study for possible superconducting phases in SrTiO$_3$ heterostructure will be presented elsewhere.

%

Throughout this work, we have stayed away from the immediate neighborhood of the quantum phase transition point,
where long-wavelength and low-frequency fluctuations of the parity-breaking order pile up and the RPA type effective interaction used in this work is inapplicable.
The physics in the quantum critical regime is an interesting topic which is left to future study.

{\it Acknowlegement:} This work is supported by the David and Lucile Packard foundation.

\bibliographystyle{apsrev}

\widetext
\begin{center}
\textbf{\large Supplemental materials} 
\end{center}
\setcounter{equation}{0}
\setcounter{figure}{0}
\setcounter{table}{0}

\makeatletter
\renewcommand{\theequation}{S\arabic{equation}}
\renewcommand{\thefigure}{S\arabic{figure}}
\renewcommand{\thetable}{S\Roman{table}}
\renewcommand{\bibnumfmt}[1]{[S#1]}
\renewcommand{\citenumfont}[1]{S#1}

This Supplementary Material consists of two sections. In section I, within an inversion-symmetric two-orbital model,
we explicitly derive the Fermi surface form factor of a parity-breaking order, which elucidates the important role
of spin-orbit coupling. In section II,  we provide a detailed derivation of the decomposition of effective interactions into different pairing channels.

\section{I. Bilayer Rashba model}

We consider a two-layer system with intrinsic spin-orbit coupling, which is the 2D analog of the model considered by Fu and Berg for
the topological insulator Bi$_2$Se$_3$~\cite{FuBerg}. The same model applies to inversion-symmetric heterostructures of SrTiO$_3$~\cite{Nagaosa}. Importantly, the two layers in the system considered here, denoted by $\tau^z$ below, are interchanged under the inversion.
It follows from symmetry that electrons on a given plane experience an out-of-plane electric field, which is opposite for the upper and lower plane.
This local electric field generates local Rashba spin-orbit coupling for electron's motion within the plane, with opposite signs for the upper and lower planes.
In addition, there is an inter-plane tunneling, which takes the off-diagonal form $m \tau^x$.
Based on symmetry considerations, the Hamiltonian for such a bilayer Rashba system, up to first order in $\bk$, is given by:
\beq
H=\int \frac{d\bk}{(2\pi)^2} \psi_{\bk}^+ [m\tau^x + v(k^x  s^y - k^y  s^x) \tau^z-\mu]\psi_{\bk}.
\eeq
where $\mu$ is chemical potential, $\tau^i$ and $s^i$ are Pauli matrices in the layer and spin space, correspondingly. Without loss of generality, we assume the chemical potential lies in the conduction band.

Since the superconducting gap is much less than chemical potential,  it is convenient to consider the conduction band only, and completely ignore valence band. To do that, we need to expand creation and annihilation operators $\psi^+_{\bk}, \, \psi_{\bk}$ in terms of the eigenstates of the conduction and valence bands, and then simply omit the contribution of the valence band. This expansion takes the form

\beq
\psi_{\bk} = \sum_{\alpha i} a_{\bk\alpha}^i c_{\bk \alpha}^i, \label{Sc}
\eeq
where $i= c, v$ labels conduction/ valence band, and $\alpha=1,2$ denote band eigenstates in the MCBB.
In this basis, eigenvectors $a_{\bk \alpha}^c$ corresponding to the states of the conduction band, which is the two-orbital analog of Bloch wavefunctions in the continuum,  have the form
\beq
a_{\bk 1}^c=\left( \begin{array}{c} \beta_+\\ i\hat k_+ \beta_- \\ \beta_+ \\ -i\hat k_+ \beta_- \end{array}   \right), \quad a_{\bk 2}^c=\left( \begin{array}{c}  -i \hat k_-\beta_- \\ \beta_+ \\ i\hat k_-\beta_-\\ \beta_+    \end{array}   \right), \label{SMCBB}
\eeq
where $\hat k_{\pm} = \hat k_x\pm i \hat k_y$, and $\beta_{\pm}=(1/2)\sqrt{1\pm (m/\mu)}$, $\mu=\sqrt{m^2+v^2 \kf^2}$. Mathematically, the mapping onto conducting band simply implies that in (\ref{Sc}) we keep $a_{\bk\alpha}^c$ only, and omit terms with $a_{\bk \alpha}^v$.

Now, we consider an inversion-symmetry-breaking charge order: an electron density imbalance on the two layers, $(n_1-n_2)(\br)=\psi^+(\br)\tau^z \psi(\br)$, which does not involve electron's spin.
After the mapping onto the conduction band, this expression can be rewritten in the form
\beq
\psi^+(\br)\tau_z \psi(\br)\to\frac {v\kf}{2\mu}\sum_{\bk, \bp} \epsilon_{ij}c_{\bk \alpha}^+ (\hat k+\hat p)^i \sigma^j_{\alpha\beta} c_{\bp\beta}e^{-i\br \cdot (\bk-\bp)},
\eeq
which exactly coincides with the Fourier transform of $\tilde \Gamma_2$ (see Table \ref{Table1}). Now, we see explicitly this form factor is pseudo-spin dependent.
This derivation illustrates how inversion-symmetry-breaking charge order in spin-orbit-coupled systems couples to the pseudospin degree of freedom on the Fermi surface.

\begin{table}[t]
\begin{center}
\begin{tabularx}{0.45\textwidth}{| X |X|}
    \hline 2-band model & 1-band model \\ \hline
    $ \tau^y s^z $ & $\tilde F_1(\bk)  =  \hat k^x \sigma^x + \hat k^y \sigma^y $\\ \hline
    $\tau^z$  & $\tilde F_2(\bk)=\hat k^x \sigma^y -\hat k^y \sigma^x $\\ \hline
    $ \tau^y s^x $ & $\tilde F_3^{x}(\bk)= \hat k^x \sigma^z $\\
   $ \tau^y s^y$  & $\tilde F_3^{y}(\bk)= \hat k^y \sigma^z $ \\
    \hline
\end{tabularx}
\end{center}
\caption{The correspondence between pairing orders in 2-band model considered in \cite{FuBerg} and 1-band model presented in this Letter.}
\label{TableS2}
\end{table}

It is also instructive to present the correspondence between p-wave pairing order parameters in single-band model and in 2-band model considered in \cite{FuBerg, nagaosa, yip}. We demonstrate the derivation of the correspondence on the example of the two-component order parameter, $\hat \Delta^i_{\bk} = \psi_{\bk \alpha}(i \tau^y s^i s^y)_{\alpha\beta}   \psi_{-\bk \beta}$. To establish the correspondence we, again, perform mapping onto conduction band using MCBB, Eq. (\ref{SMCBB}). The result of this mapping reads as follows:

\beq
\Delta^i_{\bk} = \psi_{\bk \alpha}(i \tau^y s^i s^y)_{\alpha\beta}   \psi_{-\bk \beta} \to \pm \frac{v \kf}{\mu} \hat k^i [c_{\bk1} c_{-\bk 2} + c_{\bk 2} c_{-\bk 1}]= \pm \frac{v \kf}{\mu} c_{\bk \alpha}(i\hat k^i\sigma^z \sigma^y)_{\alpha\beta} c_{-\bk \beta},
\eeq
where $+$ $(-)$ sign corresponds to $i=x$ $(i=y)$. The correspondence between other pairing orders can be obtained  analogously. The result is shown in Table \ref{TableS2}.

\section{II. Decomposition of the effective interaction into superconducting channels}

Here we show the detailed derivation of Eq. (\ref{dec}) and Table \ref{Table2}. We start with the effective interaction (\ref{vertex}) and use the approximation $V(\bk\pm\bk')=V_0$. We consider all possible types of interactions, i.e.  gyrotropic, ferroelectric and multipolar, in both three and two dimensions.

\subsection{A. Rotationally invariant interaction}

We start with the case of rotationally invariant interaction. We demonstrate the details of the calculation using gyrotropic interaction in 3D as an example, and then present the results for all other types of interaction in 3D and 2D.

The task is to decompose the effective interaction into different p-wave superconducting channels, classified in Table \ref{Table1}, and, possibly, s-wave channel. Mathematically, it means that interaction matrix elements given by Eq. (\ref{vertex}) need to be rewritten in the form:

\begin{multline}
V_{\alpha \beta\gamma \delta}(\bk,\bk')=  \frac{V_0}4\left(a_0 (i\sigma^y)_{\alpha \beta} (i\sigma^y)^+_{\gamma \delta} +a_1 (iF_1(\bk)\sigma^y)_{\alpha \beta} (iF_1(\bk')\sigma^y)^+_{\gamma \delta} + a_2 (iF_2^i(\bk)\sigma^y)_{\alpha \beta} (iF^i_2(\bk')\sigma^y)^+_{\gamma \delta}+ \right. \\ \left. +a_3 (iF_3^{ij}(\bk)\sigma^y)_{\alpha \beta} (iF^{ji}_3(\bk')\sigma^y)^+_{\gamma \delta}  \right) , \label{SV1}
\end{multline}
where summation over repeated indices $i,j=x,y,z$ is implied. All $F_i$ have the same momentum and spin dependence as $\Gamma_i$ in Table \ref{Table1}, but denote pairing in particle-particle channels.

For the case of gyrotropic interaction, the Hamiltonian has the form

\beq
H^{\rm g}_{\rm eff}= \frac{V_0}4  \sum_{\bk, \bk', \bq} a^+_{\bk\alpha} a^+_{\bk'-\bq \beta} a_{\bk' \gamma} a_{\bk-\bq \delta} \left(2\hat \bk-\hat \bq\right) \cdot \boldsymbol{\sigma}_{\alpha\delta}\left(2\hat \bk' - \hat \bq \right)\cdot \boldsymbol{\sigma}_{\beta \gamma}.
\eeq
Restricting this Hamiltonian to the Cooper channel only, we obtain the pairing Hamiltonian (\ref{Hp}) with the vertex

\beq
V_{\alpha \beta\gamma \delta}(\bk,\bk')=-\frac{V_0}8\left( (\hat \bk + \hat\bk'  )\cdot \boldsymbol{\sigma}_{\alpha\delta} (\hat \bk + \hat \bk')\cdot\boldsymbol{\sigma}_{\beta \gamma} - (\hat\bk-\hat\bk')\cdot \boldsymbol{\sigma}_{\alpha\gamma} (\hat \bk - \hat \bk') \cdot \boldsymbol{\sigma}_{\beta \delta} \right).
\eeq
This interaction corresponds to the choice $\vec d_{\bk}=\hat \bk$ in Eq. (\ref{vertex}).

As we pointed out earlier, this vertex consists of two different terms. The first one contains the product of components of the same momenta:

\beq
V^e_{\alpha \beta \gamma \delta} (\bk, \bk')=-\frac{V_0}8\sum_{i,j}(\hat k^i\hat k^j+\hat k'^i\hat k'^j)(\sigma^i_{\alpha\delta}\sigma^j_{\beta \gamma}- \sigma^i_{\alpha\gamma}\sigma^j_{\beta\delta}).
\eeq
This term describes s-wave pairing and corresponds to the first term in Eq. (\ref{SV1}). Indeed, using the identity

\beq
\sigma^i_{\alpha \delta} \sigma^j_{\beta\gamma} + \sigma^i_{\beta\gamma} \sigma^j_{\alpha\delta} - \sigma^i_{\alpha\gamma}\sigma^j_{\beta\delta} - \sigma^i_{\beta\delta} \sigma^j_{\alpha\gamma} = -2(i\sigma^y)_{\alpha\beta} (i\sigma^y)^+_{\gamma\delta} \delta^{ij}
\eeq
we find

\beq
V^e_{\alpha \beta \gamma \delta} (\bk, \bk')=\frac {V_0}4 (i\sigma^y)_{\alpha\beta}(i\sigma^y)^+_{\gamma \delta},
\eeq
therefore, $a_0=1$, and we have attraction in the s-wave pairing channel.

Next, the second term in the vertex is the bilinear function of different momenta and has the form

\beq
V^o_{\alpha \beta \gamma \delta} (\bk, \bk')=-\frac{V_0}8\sum_{i,j}(\hat k^i\hat k'^j+\hat k'^i\hat k^j)(\sigma^i_{\alpha\delta}\sigma^j_{\beta \gamma}+ \sigma^i_{\alpha\gamma}\sigma^j_{\beta\delta}).
\eeq
This term is responsible for the p-wave pairing, and need to be decomposed into p-wave superconducting channels, listed in Table \ref{Table1}. After some algebra, we find

\beq
V^o_{\alpha \beta \gamma \delta} (\bk, \bk')= \frac {V_0}4 \left\{ (i(\bk \cdot \boldsymbol{\sigma})\sigma^y)_{\alpha\beta} (i(\bk' \cdot \boldsymbol{\sigma})\sigma^y)^+_{\gamma\delta} -  (i[\bk \times \boldsymbol{\sigma}]\sigma^y)_{\alpha\beta} (i[\bk' \times \boldsymbol{\sigma}]\sigma^y)^+_{\gamma\delta} \right\}.
\eeq
After summation, $V=V^e+V^o$, we end up with the expression for the interaction vertex:

\beq
V_{\alpha \beta \gamma \delta} (\bk, \bk')= \frac {V_0}4 \left\{ (i\sigma^y)_{\alpha\beta}(i\sigma^y)^+_{\gamma \delta} + (i(\bk \cdot \boldsymbol{\sigma})\sigma^y)_{\alpha\beta} (i(\bk' \cdot \boldsymbol{\sigma})\sigma^y)^+_{\gamma\delta} -  (i[\bk \times \boldsymbol{\sigma}]\sigma^y)_{\alpha\beta} (i[\bk' \times \boldsymbol{\sigma}]\sigma^y)^+_{\gamma\delta} \right\}.
\eeq
It has the form of Eq. (\ref{SV1}) with coefficients $a_0=a_1=1, \, a_2=-1, \, a_3=0$. These are exactly the coefficients in the first row of Table \ref{Table2}.

We see that we have superconducting instabilities in s-wave and gyrotropic p-wave channels. As we mentioned above, they have identical transition temperatures, $T_c\sim \omega_0 \exp (-1/\nu |V_0|),$ where $\omega_0$ is some low-energy cutoff, and $\nu$ is the density of states at Fermi level. The detailed derivation of the transition temperatures for the case of unconventional superconductors can be found, for example, in \cite{SigristUeda,Mineev}.

All the cases of other possible interactions in three and two dimensions can be analyzed absolutely analogously. The results of the decomposition into different superconducting channels are gathered in Table \ref{Table2}.

\subsection{B. Multipolar interaction with reduced rotational symmetry}

Here we demonstrate that multipolar interaction in crystals with reduced rotational symmetry leads to the attraction in the correspondent particle-particle channel in both two and three dimensions. Moreover, in two dimensions, the critical temperature for the multipolar channels is the same as for s-wave channel.

We consider the 3D case first. We perform the same analysis as in the previous section, but now we consider diagonal and off-diagonal parts of the multipolar interaction separately. Specifically, we consider Hamiltonians

\beq
H^{\rm d}_{\rm eff}= \frac{V_0}4  \sum_{\bk, \bk', \bq} a^+_{\bk\alpha} a^+_{\bk'-\bq \beta} a_{\bk' \gamma} a_{\bk-\bq \delta} \sum_i (\Gamma^{ii}_3(2\bk-\bq))_{\alpha\delta} (\Gamma^{ii}_3(2\bk'-\bq))_{\beta \gamma}, \label{SH3d}
\eeq

\beq
H^{\rm o}_{\rm eff}= \frac{V_0}4  \sum_{\bk, \bk', \bq} a^+_{\bk\alpha} a^+_{\bk'-\bq \beta} a_{\bk' \gamma} a_{\bk-\bq \delta} \sum_{i\ne j} (\Gamma^{ij}_3(2\bk-\bq))_{\alpha\delta} (\Gamma^{ji}_3(2\bk'-\bq))_{\beta \gamma},\label{SH3o}
\eeq
where $\Gamma_3^{ij}$ are defined in Table \ref{Table1}. The whole analysis is similar to what we did before, with the only difference that now we need to consider diagonal and off-diagonal pairing channels separately:

\begin{multline}
V_{\alpha \beta\gamma \delta}(\bk,\bk')=  \frac{V_0}4\left(a_0 (i\sigma^y)_{\alpha \beta} (i\sigma^y)^+_{\gamma \delta} +a_1 (iF_1(\bk)\sigma^y)_{\alpha \beta} (iF_1(\bk')\sigma^y)^+_{\gamma \delta} + a_2 (iF_2^i(\bk)\sigma^y)_{\alpha \beta} (iF^i_2(\bk')\sigma^y)^+_{\gamma \delta}+ \right. \\ \left. +a_3^d \sum_i (iF_3^{ii}(\bk)\sigma^y)_{\alpha \beta} (iF^{ii}_3(\bk')\sigma^y)^+_{\gamma \delta}  + a_3^o \sum_{i\ne j}  (iF_3^{ij}(\bk)\sigma^y)_{\alpha \beta} (iF^{ji}_3(\bk')\sigma^y)^+_{\gamma \delta}    \right).\label{SV2}
\end{multline}
The results of the decomposition into different channels are presented in Table \ref{TableS1}.

\begin{table}[t]
\begin{center}
\begin{tabularx}{0.5\textwidth}{| l | X | X | X | X | X |}
    \hline \text{type of interaction} & $a_0$ & $a_1$ & $a_2$ & $a_3^d$ & $a_3^o$ \\  \hline
 $\left(Q_3^{ij}(\bq)Q_3^{ji}(-\bq)\right)^d$ & $8/3$ & 0 & $-2/3$& $1$ & $-1$ \\   \hline
 $\left(Q_3^{ij}(\bq)Q_3^{ji}(-\bq)\right)^o$ & $4$ & 0 & $-1$& $-3/2$ & $1/2$\\
    \hline
\end{tabularx}
\end{center}

\begin{center}
\begin{tabularx}{0.5\textwidth}{| l | X | X | X | X | X | X |}
    \hline \text{type of interaction} & $a_0$ & $a_1$ & $a_2$ & $a_3$ & $a_4^d$ & $a_4^o$ \\ \hline
 $\tilde Q_4^{xx}(\bq) \tilde Q_4^{xx}(-\bq)$ & 1 & 0 & 0 & -1 & 1 & -1
  \\ \hline
 $\tilde Q_4^{yx}(\bq) \tilde Q_4^{xy}(-\bq) $ & 1 & 0 & 0 & -1 & -1 & 1
  \\
    \hline
\end{tabularx}
\end{center}

\caption{Decomposition of the multipolar interaction with reduced rotational symmetry into pairing channels. Diagonal and off-diagonal parts of the interaction in 3D are defined as $\left(Q_3^{ij}(\bq)Q_3^{ji}(-\bq)\right)^d = \sum_i Q_3^{ii}(\bq)Q_3^{ii}(-\bq)$,  $\left(Q_3^{ij}(\bq)Q_3^{ji}(-\bq)\right)^o = \sum_{i\ne j} Q_3^{ij}(\bq)Q_3^{ji}(-\bq)$ correspondingly, see Eqn. (\ref{SH3d}), (\ref{SH3o}).}
\label{TableS1}
\end{table}

We see that, if considered separately, diagonal and off-diagonal parts of the multipolar interaction also lead to the attraction in the correspondent pairing channel. However, the interaction strength is smaller (though comparable)  than that for the s-wave channel. As a result, the transition temperature for the s-wave channel is much higher.

As we already mentioned, $\Gamma_3^{ij}$ is a traceless symmetric tensor, so it has only 5 independent components. That is why the diagonal part of $\Gamma_3^{ij}$ forms duplet $(\Gamma_3^1, \,  \Gamma_3^2)$ with

\beq
\Gamma_3^1=\Gamma_3^{xx} - \Gamma_3^{yy}, \qquad \Gamma_3^2= 2\Gamma_3^{zz} - \Gamma_3^{xx} - \Gamma_3^{yy},
\eeq
rather than triplet $(\Gamma_3^{xx}, \, \Gamma_3^{yy}, \, \Gamma_3^{zz}).$ Hopefully, the diagonal part of the interaction can easily be rewritten in terms of $(\Gamma_3^1, \, \Gamma_3^2)$ in both particle-hole and particle-particle channels:

\begin{multline}
(\Gamma_3^{xx}(\bk))_{\alpha\delta} (\Gamma_3^{xx} (\bk'))_{\beta \gamma} + (\Gamma_3^{yy}(\bk))_{\alpha\delta} (\Gamma_3^{yy} (\bk'))_{\beta \gamma} + (\Gamma_3^{zz}(\bk))_{\alpha\delta} (\Gamma_3^{zz} (\bk'))_{\beta \gamma} = \\ = 2 (\Gamma_3^{1}(\bk))_{\alpha\delta} (\Gamma_3^{1} (\bk'))_{\beta \gamma} + \frac23 (\Gamma_3^{2}(\bk))_{\alpha\delta} (\Gamma_3^{2} (\bk'))_{\beta \gamma} ,
\end{multline}

\begin{multline}
(i\Gamma_3^{xx}(\bk) \sigma^y)_{\alpha\beta} (i\Gamma_3^{xx} (\bk') \sigma^y)^+_{\gamma \delta} + (i\Gamma_3^{yy}(\bk) \sigma^y)_{\alpha\beta} (i\Gamma_3^{yy} (\bk') \sigma^y)^+_{\gamma \delta} + (i\Gamma_3^{zz}(\bk) \sigma^y)_{\alpha\beta} (i\Gamma_3^{zz} (\bk') \sigma^y)^+_{\gamma \delta} = \\ = 2 (i\Gamma_3^{1}(\bk) \sigma^y)_{\alpha\beta} (i\Gamma_3^{1} (\bk') \sigma^y)^+_{\gamma \delta} + \frac 23 (i\Gamma_3^{2}(\bk) \sigma^y)_{\alpha\beta} (i\Gamma_3^{2} (\bk') \sigma^y)^+_{\gamma \delta}.
\end{multline}

Finally, we consider the multipolar interaction with broken rotational symmetry in 2D. Unlike 3D case, $\tilde \Gamma_4$ splits now into two one-dimensional representations with form factor $\tilde \Gamma_4^{xx}(\bk)= k^x\sigma^x- k^y \sigma^y$ and $\tilde \Gamma_4^{xy} (\bk)= k^x\sigma^y + k^y \sigma^x$. Consequently, diagonal and off-diagonal parts of interaction are given now by

\beq
H^{\rm d}_{\rm eff}= \frac{V_0}4  \sum_{\bk, \bk', \bq} a^+_{\bk\alpha} a^+_{\bk'-\bq \beta} a_{\bk' \gamma} a_{\bk-\bq \delta} (\tilde\Gamma^{xx}_4(2\bk-\bq))_{\alpha\delta} (\tilde\Gamma^{xx}_4(2\bk'-\bq))_{\beta \gamma}, \label{SH2d}
\eeq

\beq
H^{\rm o}_{\rm eff}= \frac{V_0}4  \sum_{\bk, \bk', \bq} a^+_{\bk\alpha} a^+_{\bk'-\bq \beta} a_{\bk' \gamma} a_{\bk-\bq \delta} (\tilde\Gamma^{xy}_4(2\bk-\bq))_{\alpha\delta} (\tilde\Gamma^{yx}_4(2\bk'-\bq))_{\beta \gamma}.\label{SH2o}
\eeq
We emphasize that there is no any summation over different components of $\tilde \Gamma_4$ here. Indeed, due to the equalities $\tilde\Gamma^{xx}_4(\bk)= - \tilde\Gamma^{yy}_4(\bk)$ and $\tilde\Gamma^{xy}_4(\bk)=  \tilde\Gamma^{yx}_4(\bk),$ this summation would only lead to the overall factor 2 in front of the interaction Hamiltonian. Analogously, we do not double count different components when decompose the effective interaction into different pairing channels.

The results of the decomposition are again gathered in Table \ref{TableS1}. In contrast to the 3D case, s-wave channel and corresponding multipolar channels now have the same interaction strength, and, as a result, the same critical temperature $T_c\sim \omega_0 \exp (-1/\nu |V_0|).$

In order to obtain the last row of Table \ref{Table2}, one need to sum up diagonal and off-diagonal contributions from Table \ref{TableS1} and to multiply it by 2. This factor 2 comes exactly from the double counting of the components of $\tilde \Gamma_4$ that we mentioned above.

\bibliographystyle{apsrev}

\begin{thebibliography}{10}
\bibitem{footnote}
The most general form factor of the order parameter of He-3 in BW phase is a rank 2 tensor $k_i s_j$, because in the absence of spin-orbit coupling
the spin and momentum are free to rotate separately.
\end{thebibliography}

\end{document}